\documentstyle[amssymb,preprint,aps]{revtex}
\begin{document}
\draft
\title{Andreev-Fano Effect in a Hybrid\\
Normal-Metal / Superconductor Interferometer}
\author{{\ Yu Zhu}$^1${, Qing-feng Sun}$^2${, and }Tsung-han Lin$^{1*}$}
\address{{\it State Key Laboratory for Mesoscopic Physics and }\\
{\it Department of Physics, Peking University,}{\small \ }{\it Beijing}\\
100871, China$^1$}
\address{{\it Center for the Physics of Materials and}\\
{\it Department of Physics, McGill University, Montreal,}\\
{\it PQ, Canada H3A 2T8}$^2$}
\date{}
\maketitle

\begin{abstract}
We report on a new type of Fano effect, named as Andreev-Fano effect, in a
hybrid normal-metal / superconductor (N/S) interferometer embedded with a
quantum dot. Compared with the conventional Fano effect, Andreev-Fano effect
has some new features related to the characteristics of Andreev reflection.
In the linear response regime, the line shape is the square of the
conventional Fano shape; while in the nonlinear transport, a sharp resonant
structure is superposed on an expanded interference pattern, qualitatively
different from the conventional Fano effect. The phase dependence of the
hybrid N/S interferometer is also distinguished from those of all-N or all-S
interferometers.
\end{abstract}

%\vskip 0.4in

PACS numbers: 74.50.+r, 73.63.Kv, 85.35.Ds, 73.23.-b

\baselineskip 20pt %\baselineskip 12pt
\newpage

When a resonant channel interferes with a nonresonant channel, the line
shape of the resonance will change into an asymmetric one, sometimes even
become to an antiresonance. This phenomenon is known as Fano effect \cite
{Fano}, which has been recognized in a large variety of systems, such as
atomic photoionization, electron and neutron scattering, Raman scattering,
and photoabsorption in quantum well structures, etc. Recently, the effect is
observed in the electron transport through mesoscopic systems. Several
groups reported Fano resonance in spectroscopic measurements of a single
magnetic adatom absorbed on the surface of normal metal, using scanning
tunneling microscope \cite{Fexp1,Fexp2,Fexp3,Fexp4}. The results can be well
understood by generalizing the noninteracting resonant channel in the
original Fano problem to the many-body Kondo resonance \cite
{Fthy1,Fthy2,Fthy3}. Meanwhile, Fano resonance is also reported in the
conductance through a single electron transistor fabricated in semiconductor 
\cite{Fexp5}, which has the advantage that the key parameters are
experimentally controllable.

Another active field in recent years is the so called ``mesoscopic
superconductivity'', which has been greatly advanced with the progress of
nanofabrication technology \cite{MS2000}. Adding superconducting materials
to the conventional mesoscopic systems changes the coherent transport
through the hybrid system significantly. At the interface of normal-metal
(N) and superconductor (S), a two-particle process called Andreev reflection
(AR) plays an essential role in the subgap conductance, in which an electron
is reflected as a hole in the N side, and a Cooper pair is created in the S
side \cite{Andreev}. Many new effects involving AR\ are investigated both
experimentally and theoretically (for a recent review see \cite{review}).

What will occur if one of the electrodes in the mesoscopic Fano system is
replaced by a superconductor? Let us consider a N/S hybrid interferometer,
one arm contains a point contact with tunable conductance, the other arm is
embedded with a quantum dot (QD) with resonant levels. The proposed
structure is schematically shown in Fig.1, and hereafter is referred to as
N-(I,QD)-S. The conductance through the path N-I-S\ provides a nonresonant
channel, and the transport properties have been studied in \cite{BTK,Happ}.
While the conductance through the path N-QD-S\ provides a resonant channel,
and the transport properties have been studied in \cite{Beenakker,ar1}. The
interplay of the two channels in N-(I,QD)-S are double folded: interference
between resonant AR and nonresonant AR, and opening up a new conducting
channel, named as cross AR. Due to the interference among these AR
processes, we find a new type of Fano effect, named as Andreev-Fano effect,
which has distinct features compared with the conventional Fano effect. In
the linear response regime, the line shape is the square of the conventional
Fano shape; while in the nonlinear transport, a sharp resonant structure is
superposed on an expanded interference pattern. The phase dependence of the
hybrid N/S interferometer is also distinguished from those of all-N or all-S
interferometers.

The N-(I,QD)-S structure is modelled by the Hamiltonian 
\begin{equation}
H=H_L+H_R+H_{dot}+H_T\;\;,
\end{equation}
where $H_L=\sum_{k\sigma }\varepsilon _{Lk}a_{Lk\sigma }^{\dagger
}a_{Lk\sigma }$ is for the left N electrode, $H_R=\sum_{k\sigma }\varepsilon
_{Rk}a_{Rk\sigma }^{\dagger }a_{Rk\sigma }+\sum_k(\Delta a_{Rk\uparrow
}^{\dagger }a_{R-k\downarrow }^{\dagger }+H.c.)$ is for the right S
electrode, $H_{dot}=\sum_\sigma E_0c_\sigma ^{\dagger }c_\sigma $ is for the
QD embedded in one arm of the interferometer, and $H_T=t_L\sum_{k\sigma
}a_{Lk\sigma }^{\dagger }c_\sigma +t_R\sum_{k\sigma }a_{Rk\sigma }^{\dagger
}c_\sigma +w\sum_{kk^{\prime }}a_{Lk\sigma }^{\dagger }a_{Rk^{\prime }\sigma
}+H.c.$ is for the phase coherent tunneling within the interferometer.

To proceed, we introduce the following Green function matrix 
\begin{equation}
{\bf G\equiv \langle \langle }\left( 
\begin{array}{l}
C \\ 
A_L \\ 
A_R
\end{array}
\right) |\left( 
\begin{array}{lll}
C^{\dagger } & A_L^{\dagger } & A_R^{\dagger }
\end{array}
\right) {\bf \rangle \rangle =}\left( 
\begin{array}{lll}
G_{DD} & G_{DL} & G_{DR} \\ 
G_{LD} & G_{LL} & G_{LR} \\ 
G_{RD} & G_{RL} & G_{RR}
\end{array}
\right) {\bf \;\;,}
\end{equation}
in which 
\begin{equation}
C=\left( 
\begin{array}{l}
c_{\uparrow } \\ 
c_{\downarrow }^{\dagger }
\end{array}
\right) ,\;A_L=\left( 
\begin{array}{l}
\sum_ka_{Lk\uparrow } \\ 
\sum_ka_{L-k\downarrow }^{\dagger }
\end{array}
\right) ,\;A_R=\left( 
\begin{array}{l}
\sum_ka_{Rk\uparrow } \\ 
\sum_ka_{R-k\downarrow }^{\dagger }
\end{array}
\right) \;\;.
\end{equation}
One can write down the Dyson equation of the system as ${\bf G=g+g\Sigma G}$%
, with ${\bf g}$ being the decoupled Green function in the limit of $%
H_T\rightarrow 0$, and ${\bf \Sigma }$ being the self-energy arise from the
coherent tunneling. The corresponding Dyson equation for ${\bf G}^r$ and
Keldysh equation for ${\bf G}^{<}$ read 
\begin{eqnarray}
{\bf G}^r &=&\left( {\bf g}^{r^{-1}}-{\bf \Sigma }\right) ^{-1}\;\;\;\;\;, \\
{\bf G}^{<} &=&{\bf G}^r{\bf g}^{r^{-1}}{\bf g}^{<}{\bf g}^{a^{-1}}{\bf G}%
^a\;\;,
\end{eqnarray}
in which ${\bf \Sigma =\Sigma }^a{\bf =\Sigma }^r$ is the matrix of
tunneling elements.

The current flowing out of the electrode $\beta $ can be expressed in term
of these Green functions, 
\begin{equation}
I_{\beta \sigma }=\frac e\hbar 2%
%TCIMACRO{\func{Re}}
%BeginExpansion
\mathop{\rm Re}%
%EndExpansion
\int \frac{d\omega }{2\pi }Tr\left[ {\bf Q}_{\beta \sigma }{\bf \Sigma G}%
^{<}\right] ,\;\;(\beta =L,R)
\end{equation}
in which 
\begin{eqnarray}
{\bf Q}_{L\sigma } &=&\left( 
\begin{array}{lll}
0 &  &  \\ 
& q_\sigma  &  \\ 
&  & 0
\end{array}
\right) ,\;{\bf Q}_{R\sigma }=\left( 
\begin{array}{lll}
0 &  &  \\ 
& 0 &  \\ 
&  & q_\sigma 
\end{array}
\right) , \\
q_{\uparrow } &=&\left( 
\begin{array}{cc}
1 & 0 \\ 
0 & 0
\end{array}
\right) ,\;q_{\downarrow }=\left( 
\begin{array}{cc}
0 & 0 \\ 
0 & -1
\end{array}
\right) ,\;\;
\end{eqnarray}
are matrices with element $+1$ for electron and $-1$ for hole. For the
non-spin-polarized case considered in this paper, the total current through
the interferometer is $I=I_{L\uparrow }+I_{L\downarrow }=2I_{L\uparrow }$.
The current can be separated into two parts: the AR current $I_A$ and the
conventional tunneling current $I_B$, 
\begin{eqnarray}
I_A &=&\frac{2e}\hbar \int \frac{d\omega }{2\pi }T_A(\omega )\left[ f(\omega
-eV)-f(\omega +eV)\right] \;\;, \\
I_B &=&\frac{2e}\hbar \int \frac{d\omega }{2\pi }T_B(\omega )\left[ f(\omega
-eV)-f(\omega )\right] \;\;\;\;\;\;\;\;.
\end{eqnarray}
It can be shown that $T_B(\omega )$ contains a factor $\frac{|\omega |}{%
\sqrt{\omega ^2-\Delta ^2}}\theta (|\omega |-\Delta )$, and therefore is
negligible when $\left| eV\right| <\Delta $. $T_A(\omega )$ is derived as 
\begin{equation}
T_A(\omega )=2%
%TCIMACRO{\func{Re}}
%BeginExpansion
\mathop{\rm Re}%
%EndExpansion
Tr\left[ {\bf Q}_{L\uparrow }{\bf \Sigma G}^r{\bf Q}_{L\downarrow }{\bf (g}%
^{r^{-1}}{\bf -g}^{a^{-1}}{\bf )G}^a\right] \;\;.
\end{equation}
Since we are concerning the subgap conductance where AR\ dominates, we take
the limit $\Delta \rightarrow \infty $ in the evaluation of $T_A(\omega )$.
After some algebra, $T_A(\omega )$ can be explicitly obtained as 
\begin{equation}
T_A(\omega )=\frac{\left| \frac 12LR-2\sqrt{xLR}E_0e^{\text{i}\phi
}-2x(\omega ^2-E_0^2)e^{\text{i}2\phi }\right| ^2}{\left[ (1+x^2)(\omega
^2-E_0^2)-\frac{L^2+R^2}4+2x\sqrt{xLR}E_0\cos \phi -xLR\cos ^2\phi \right]
^2+\left[ (L+xR)\omega \right] ^2}
\end{equation}
where $L\equiv 2\pi N_L\left| t_L\right| ^2$, $R\equiv 2\pi N_R\left|
t_R\right| ^2$, and $x\equiv \pi ^2N_LN_R\left| w\right| ^2$ are the
coupling strengths among N, S, and QD, with $N_L$ and $N_R$ being the
density of states in the left and right electrodes. $\phi $ is the
Aharonov-Bohm (AB) phase induced by a magnetic flux. It can be shown that $%
0\leqslant T_A(\omega )\leqslant 1$ as required by the physical meaning of
transmission probability. Specially, if $x\rightarrow 0$, $T_A(\omega )=%
\frac 14L^2R^2/\left[ \left( \omega ^2-E_0^2-\frac{L^2+R^2}4\right)
^2+L^2\omega ^2\right] $ reproduces the transmission probability of N-QD-S;
if $L\rightarrow 0$ and $R\rightarrow 0$, $T_A(\omega )=\frac{4x^2}{(1+x^2)^2%
}$ reproduces the transmission probability of N-I-S. $\;$

At zero temperature and in the subgap regime, the total current is reduced
to $I=\frac{2e}\hbar \int_{-eV}^{eV}\frac{d\omega }{2\pi }T_A(\omega )$. The
integral can be evaluated analytically in two limits: linear response regime
where $eV\rightarrow 0$ and strong nonlinear regime where $eV\rightarrow
\infty $ \cite{remark1}. For $eV\rightarrow 0$, the conductance at zero bias
is 
\begin{equation}
G_0\equiv \lim_{V\rightarrow 0}\frac{I(V)}V=\frac{4e^2}hT_b\left[ \frac{%
\left( E_0-\frac{\sqrt{xLR}}{2x}\cos \phi \right) ^2+\frac{LR}{4x}\sin
^2\phi }{\left( E_0-\frac{x\sqrt{xLR}}{1+x^2}\cos \phi \right) ^2+\frac{xLR}{%
(1+x^2)^2}\cos ^2\phi +\frac{L^2+R^2}{4(1+x^2)}}\right] ^2\;\;,
\end{equation}
in which $T_b\equiv \frac{4x^2}{(1+x^2)^2}$ is the transmission probability
through the arm of N-I-S. For $eV\rightarrow \infty $, the net current in
high voltage limit is 
\begin{eqnarray}
I_\infty &\equiv &\lim_{V\rightarrow \infty }\left[ I(V)-\frac{4e^2}h%
T_bV\right] \\
&=&\frac{2e}\hbar \frac 2{(L+xR)(1+x^2)}\left[ xLR\sin ^2\phi -\frac{x^2}{%
(1+x^2)^2}(L+xR)^2\right.  \nonumber \\
&&+\left. \frac{\left( \frac{1-x^2}{1+x^2}\sqrt{xLR}\cos \phi E_0+\frac{x^2}{%
1+x^2}LR\cos ^2\phi +\frac{x(L^2+R^2)}{4(1+x^2)}-\frac{LR}4\right) ^2}{%
\left( E_0-\frac{x\sqrt{xLR}}{1+x^2}\cos \phi \right) ^2+\frac{xLR\cos
^2\phi }{(1+x^2)^2}+\frac{^{L^2+R^2}}{4\left( 1+x^2\right) }}\right] \;, 
\nonumber
\end{eqnarray}
in which the background current through the arm of N-I-S has been subtracted
from the total current for convergence.

Equation (12), (13), and (14) are the central results of this paper, which
describe the Andreev-Fano effect in the N-(I,QD)-S structure. Below we shall
discuss in detail the physical meaning of these results by numerical
calculation.

Fig.2 shows the curves of net conductance $G\equiv G_0-\frac{4e^2}hT_b$ at
zero bias voltage and the net current $I\equiv I(V)-\frac{4e^2}hT_bV$ at
finite voltages for vanishing AB phase, with $\sqrt{L^2+R^2}\equiv 1$ as
energy unit. For comparison, the corresponding curves in N-(I,QD)-N are also
shown in the plot, where the background conductance $\frac{4e^2}hT_b$ should
be replaced by $\frac{2e^2}hT_b^{\prime }$ with $T_b^{\prime }=\frac{4x}{%
(1+x)^2}$, and the energy unit replaced by $(L+R)/(1+x)$. The background
conductance $\frac{4e^2}hT_b$ is contributed by the transport through the
arm N-I-S, while the net conductance (current) is contributed by remaining
conducting paths and the interference among all channels. The interference
is so important in the coherent transport that the net conductance (current)
could even be negative. In the linear response regime, one can see in the
plot that the pattern of conventional Fano effect and Andreev-Fano effect
are analogous. With the increase of background conductance, the original
resonant peak becomes asymmetric, then evolves into a positive and a
negative peak, finally into an antiresonance which is the result of
destructive interference. Despite of the similarity, the line shapes are
different: the conventional Fano effect has the shape of $\frac 1{1+q^2}%
\frac{(\varepsilon +q)^2}{\varepsilon ^2+1}$ while the Andreev-Fano effect
has $\left[ \frac 1{1+q^2}\frac{(\varepsilon +q)^2}{\varepsilon ^2+1}\right]
^2$, where $\varepsilon $ is the effective resonant level and $q$ the Fano
parameter.

At finite voltages, the conventional Fano effect and the Andreev-Fano effect
are qualitatively different, which is apparent by comparing (a3) with (b2)
in Fig.2. In conventional Fano effect, the resonant structure at zero bias
voltage is pulled to a flat ridge or valley by finite voltages, while in
Andreev-Fano effect, the resonant structure near $E_0=\mu _R$ remains sharp
even in the high voltage limit, and the resonance reverses its sign from
positive to negative then return to positive with the increase of background
conductance. In addition, the sharp resonant structure is superposed on an
expanded interference pattern in the range of $eV<\left| E_0-\mu _R\right| $%
. These differences stem from the different conducting mechanisms. In
N-(I,QD)-N, electron transport is dominated by the single particle process.
The N-QD-N arm becomes conductive when the resonant level $E_0$ is within
the chemical potentials of the left and right N electrodes. Therefore, the
interference pattern is simply expanded to the ranges of $\mu _L<E_0<\mu _R$
by the finite voltages. In N-(I,QD)-S, on the contrast, electron transport
in the subgap regime is governed by AR. There are three types of AR in the
N-(I,QD)-S system (see fig.1), (a) resonant AR through the path N-QD-S, (b)
non-resonant AR through the path N-I-S, and (c) cross AR with the incident
electron coming from N-QD-S and the reflected hole going through N-I-S. The
features of Andreev-Fano effect at finite voltages can be well interpreted
with the these AR processes. When $\left| E_0-\mu _R\right| >eV$, only
process (b) is allowed, resulting in a flat background current. So the net
current $I\rightarrow 0$ after subtracting the background. When $eV>\left|
E_0-\mu _R\right| >O(1)\sqrt{L^2+R^2}$, process (b) and (c) are active,
their interference lead to a current pattern similar to that of N-(I,QD)-N.
When $\left| E_0-\mu _R\right| <O(1)\sqrt{L^2+R^2}$, the conducting channel
(c) is open. (Notice that resonant AR occurs only when the resonant level
lines up with the chemical potential of S, because both the incident
electron and the reflected hole have to pass QD through the resonant level.)
Consequently, a resonant structure is superposed on the former interference
pattern. The interference among (a), (b), and (c) makes the resonant peak
reverse its sign twice. For a quantitative analysis, let us consider $%
I_\infty $ at $\phi =0$. The current formula can be formally rewritten to a
conventional Fano shape $I_\infty =A_1+A_2\frac{(\varepsilon +q)^2}{%
\varepsilon ^2+1}$, in which 
\[
q=\frac{2x\sqrt{xLR}+\frac{1+x^2}{1-x^2}\frac{x(L^2+R^2)+LR(3x^2-1)}{2\sqrt{%
xLR}}}{\sqrt{(L+xR)^2+(R+xL)^2}} 
\]
is the effective Fano parameter of the resonant peak. Obviously, $%
q\rightarrow -\infty $ if $T_b\rightarrow 0$ , $q\rightarrow +\infty $ if $%
T_b\rightarrow 1$, and $q=0$ at some intermediate $T_b$, which are
corresponding to the sign reverse of the resonant peak.

Next, we investigate the phase dependence of the hybrid N/S interferometer.
Notice that the current has the following features: $I(\phi +2\pi )=I(\phi )$
due to the general property of AB rings, $I(\pi -\phi ,-E_0)=I(\phi ,E_0)$
due to particle and hole symmetry, and $I(-\phi )=I(\phi )$ analogous to the
phase locking effect in conventional two-terminal structures \cite{phaselock}
but here the conducting mechanism is AR. The left panel of Fig.3 shows the
curves of $G$ vs $E_0$ at zero bias voltage and $I$ vs $E_0$ at finite
voltages, with $T_b$ fixed at 0.5 and $\phi $ chosen as $0$, $\frac \pi 4$,
and $\frac \pi 2$. (Only the cases of $0\leqslant \phi \leqslant \frac \pi 2$
are shown due to the features of $I(\phi )$.) In the linear response regime,
the interference pattern evolves from asymmetric to symmetric shape when $%
\phi $ changes from $0$ to $\frac \pi 2$, but the dependence on $\phi $ is
insensitive. At finite voltages, however, the pattern exhibits a strong
dependence on $\phi $. Especially, the current platform in the high voltage
limit oscillates with $\phi $ according to $-\frac e\hbar \frac{LR}{L+xR}%
T_b^{3/2}\cos 2\phi $. The right panel of fig.3 shows the curves of $G$ vs $%
\phi $ and $I$ vs $\phi $ at selected points marked in the left curves. In
the linear response regime, the phase dependence is dominated by $\cos \phi $
or $\cos (\phi +\pi )$ component with an abrupt $\pi $ phase shift around $%
E_0=\mu _R$. While in the high voltage limit the phase dependence is
dominated by $\cos 2\phi $ component without any abrupt phase shift. At
intermediate voltages, the phase dependence is much more complicated: $\cos
\phi $ and $\cos (\phi +\pi )$ dominate in the range of $E_0-\mu _R<-eV$ and 
$E_0-\mu _R>eV$, respectively; $\cos \phi $ and $\cos 2\phi $ coexist in the
range of $-eV<E_0-\mu _R<0$; $\cos (\phi +\pi )$ and $\cos 2\phi $ coexist
in the range of $0<E_0-\mu _R<eV$; both $\cos \phi $ and $\cos (\phi +\pi )$
vanish around $E_0=\mu _R$ and only $\cos 2\phi $ dominates there. Despite
of the constraint of the phase locking, our results suggest that more
information could be extracted from the analysis of higher harmonics $\cos
2\phi $. The results share some similarity with the phase dependence of
Kondo-Fano effect studied in \cite{Fthy3}. The coexistence of $\cos \phi $
and $\cos 2\phi $ harmonics in the hybrid N/S interferometer is can be
understood as follows. Taking account of various AR processes and
considering only the direct trajectory, the transmission probability through
the interferometer is $\left| t_{1\uparrow }t_{1\downarrow }+t_{2\uparrow
}e^{\text{i}\phi }t_{2\downarrow }e^{\text{i}\phi }+t_{1\uparrow
}t_{2\downarrow }e^{\text{i}\phi }+t_{1\uparrow }t_{2\downarrow }e^{\text{i}%
\phi }\right| ^2$, in which $t_{1\sigma }$ ($t_{2\sigma }$) is the tunneling
amplitude through the upper (lower) arm. Both $\cos \phi $ and $\cos 2\phi $
emerge in the interference terms. One can see in Eq.(12) that the numerator
of $T_A(\omega )$ is just of this form, while the additional $\cos \phi $
and cos$^2\phi $ terms in the denominator can be attributed to the
trajectory with higher winding number.

To sum up, we have reported the Andreev-Fano effect in a hybrid N/S
interferometer, and found some new features compared with the conventional
Fano effect. Both the interference pattern and the phase dependence of the
hybrid N/S interferometer are different to the conventional all-N or all-S
interferometer, which originates from different conducting mechanism. We
believe that the proposed structure can be achieved with the available
technique, e.g., replacing the N-tip of scanning tunneling microscope by a
S-tip in the mesoscopic Fano system \cite{Fexp1,Fexp2,Fexp3,Fexp4}, or
modifying the structure of the Andreev interferometer in the experiment \cite
{ARmeter}, or fabricating the design in the S-2DEG hybrid systems \cite
{S/2DEG1}, etc. We are looking forward to hearing the relevant experimental
response.

The authors would like to thank Z. S. Ma for stimulating discussion. This
project was supported by NSFC\ under Grants No. 10074001 and No. 90103027,
and also by the Visiting Scholar Foundation of State Key Laboratory for
Mesoscopic Physics in Peking University.

\smallskip $^{*}$ To whom correspondence should be addressed.

%\section* {REFERENCES}

\section*{Figure Captions}

\begin{itemize}
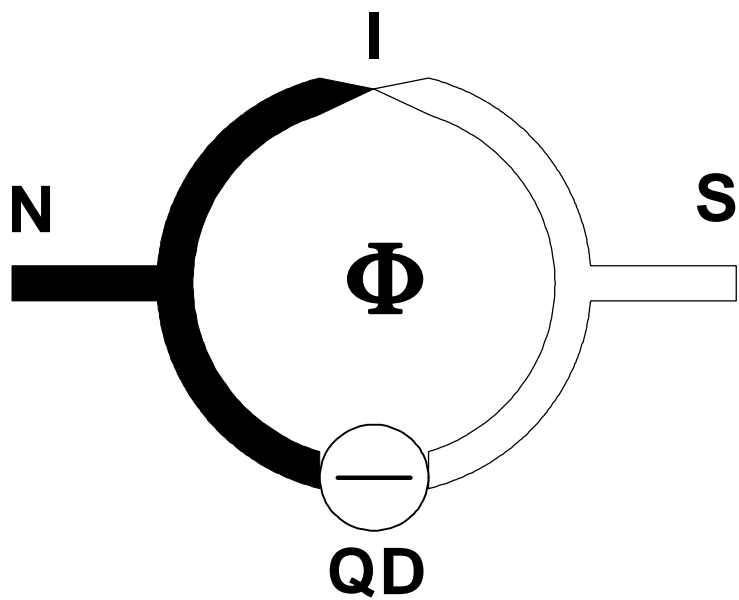
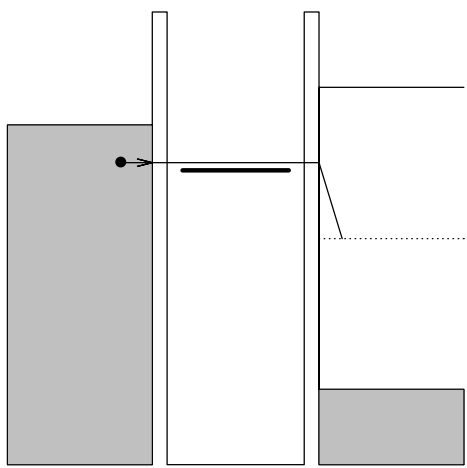
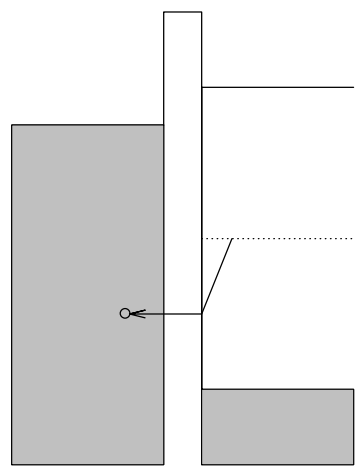
\item[{\bf Fig. 1}]  Schematic diagram of the hybrid N/S interferometer.
Below is the illustration of several AR processes: (a) is the resonant AR
through N-QD-S, (b) is the nonresonant AR through N-I-S, and (c) is the
cross AR between N-QD-S and N-I-S.

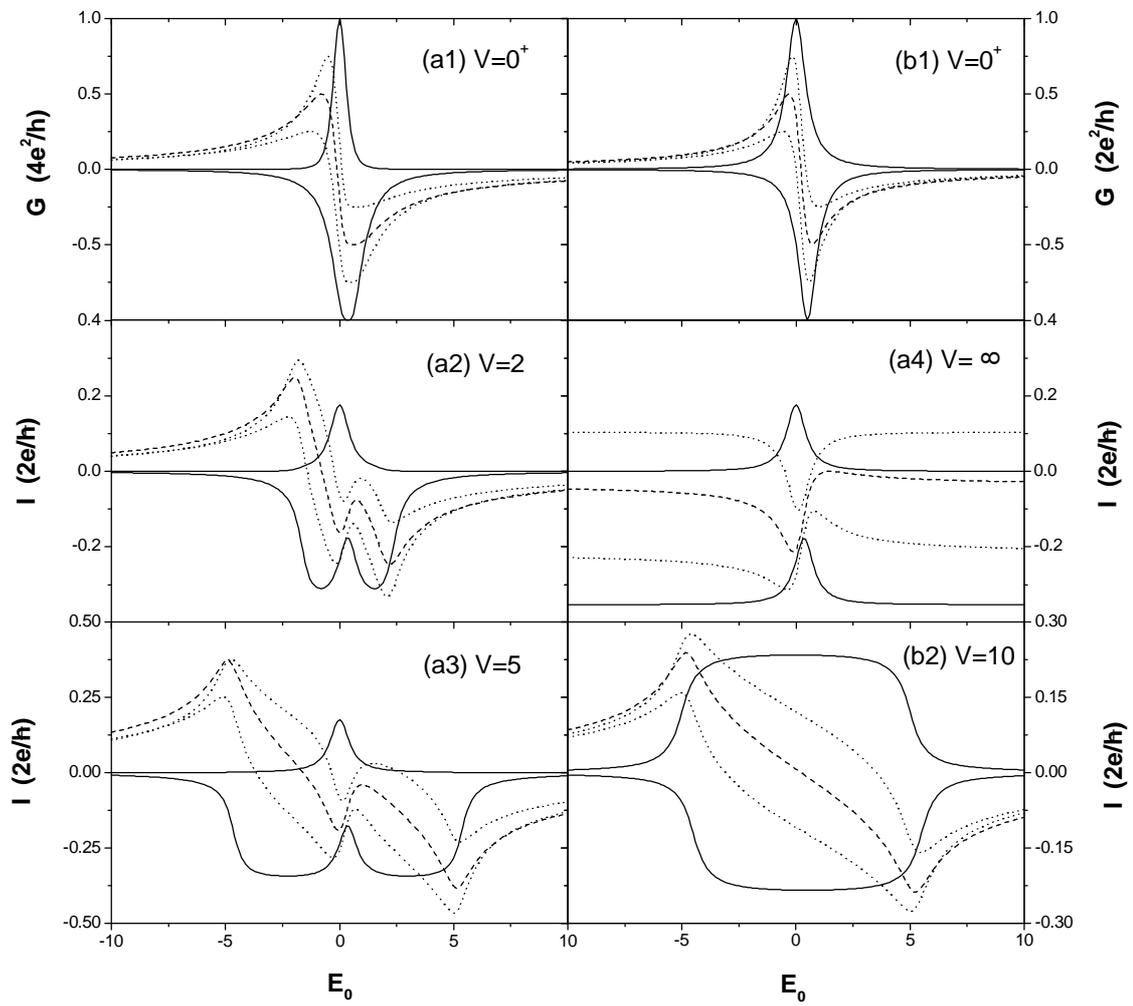
\item[{\bf Fig. 2}]  Linear and nonlinear transport though (a) N-(I,QD)-S
and (b) N-(I,QD)-N, with vanishing AB\ phase. For clarity, the background
current through the reference arm N-I-S or N-I-N has been subtracted from
the total current. The background transmission probability $T_b$ is chosen
as $0$ (solid), $\frac 14$ (dot), $\frac 12$ (dash), $\frac 34$ (dot), $1$
(solid), with the symmetric coupling strengths $L=R$.

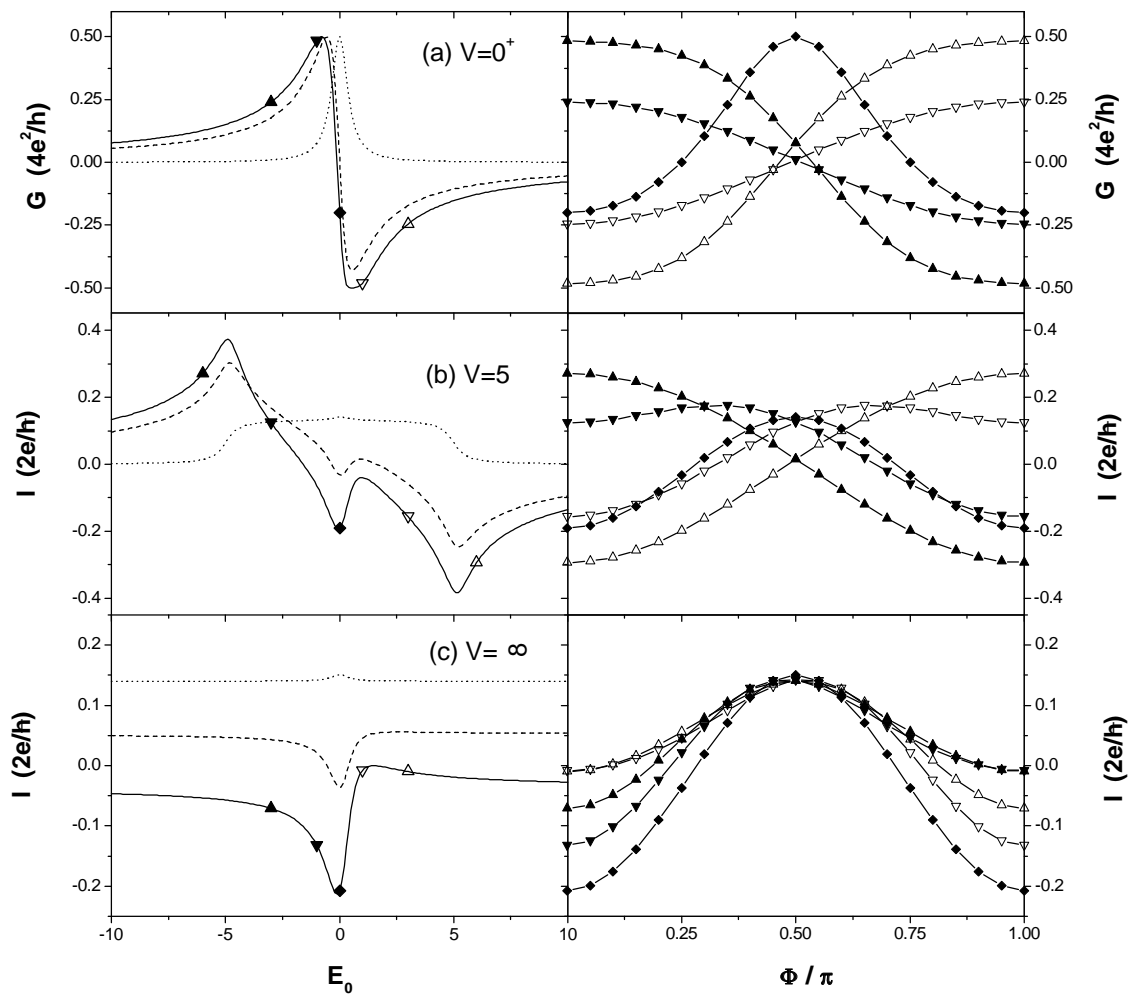
\item[{\bf Fig. 3}]  Phase dependence of the N-(I,QD)-S interferometer. The
left panel shows the curves of $G$ vs $E_0$ and $I$ vs $E_0$ for $T_b=\frac 1%
2$, with fixed AB phase $\phi =0$ (solid), $\frac \pi 4$ (dash), and $\frac %
\pi 2$ (dot). The right panel shows the curves of $G$ vs $\phi $ and $I$ vs $%
\phi $ at the selected points marked in the left curves.
\end{itemize}

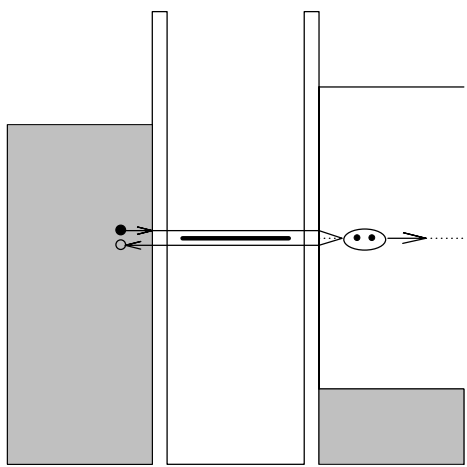
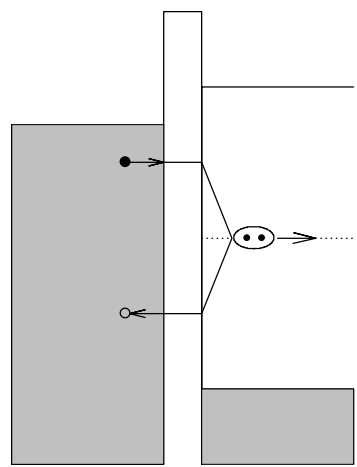
\end{document}